\documentclass[conference]{IEEEtran}

\usepackage{cite}
\usepackage{amsmath,amssymb,amsfonts}
\usepackage{algorithmic}
\usepackage{graphicx}
\usepackage{textcomp}
\usepackage{fancyhdr}
\usepackage{comment}
\usepackage{multirow}
\usepackage{tikz}

\usepackage[T1]{fontenc}
\usepackage{booktabs} 
\usepackage{soul}
\usepackage{amssymb}
\usepackage{pifont}

\usepackage[bookmarks=true,breaklinks=true,letterpaper=true,colorlinks,linkcolor=blue,citecolor=blue,urlcolor=black]{hyperref}
\newcommand{\insertFigure}[2]{
    \begin{figure}[t!]
        \centering
        \includegraphics[width=\linewidth]{figures/#1.pdf}
	\vspace{-6mm}
        \caption{\small #2}
	\vspace{-3mm}
        \label{fig:#1}
    \end{figure}
}

\newcommand{\insertWideFigure}[2]{

    \begin{figure*}[ht!]
        \centering
        \includegraphics[width=\textwidth]{figures/#1.pdf}
	\vspace{-6mm}
        \caption{\small #2}
	\vspace{-3mm}
        \label{fig:#1}
    \end{figure*}

}

\newcommand{\squishlist}{
 \begin{list}{$\bullet$}
  { \setlength{\itemsep}{0pt}
     \setlength{\parsep}{0pt}
     \setlength{\topsep}{3pt}
     \setlength{\partopsep}{0pt}
     \setlength{\leftmargin}{1.5em}
     \setlength{\labelwidth}{1em}
     \setlength{\labelsep}{0.5em} } }
     
\newcommand{\squishnums}{
 \begin{list}{$\bullets$}
  { \setlength{\itemsep}{0pt}
     \setlength{\parsep}{3pt}
     \setlength{\topsep}{3pt}
     \setlength{\partopsep}{0pt}
     \setlength{\leftmargin}{1.5em}
     \setlength{\labelwidth}{1em}
     \setlength{\labelsep}{0.5em} } }

\newcommand{\squishlisttwo}{
 \begin{list}{$\bullet$}
  { \setlength{\itemsep}{0pt}
     \setlength{\parsep}{0pt}
    \setlength{\topsep}{0pt}
    \setlength{\partopsep}{0pt}
    \setlength{\leftmargin}{2em}
    \setlength{\labelwidth}{1.5em}
    \setlength{\labelsep}{0.5em} } }

\newcommand{\squishend}{
  \end{list}  }

\newcommand{\Rav}[1]{{\color{orange}\bfseries [Raveesh::: #1]}}

\newcommand{\cmark}{\ding{51}}%
\newcommand{\xmark}{\ding{55}}%

\renewcommand{\hl}{}

\usepackage[normalem]{ulem}

\IEEEoverridecommandlockouts
\usepackage{cite}
\usepackage{amsmath,amssymb,amsfonts}
\usepackage{algorithmic}
\usepackage{graphicx}
\usepackage{textcomp}
\usepackage{xcolor}
\def\BibTeX{{\rm B\kern-.05em{\sc i\kern-.025em b}\kern-.08em
    T\kern-.1667em\lower.7ex\hbox{E}\kern-.125emX}}
\begin{document}

\title{
Understanding the Design-Space of Sparse/Dense Multiphase GNN dataflows on Spatial Accelerators
}

\author{\IEEEauthorblockN{Raveesh Garg\IEEEauthorrefmark{1}, Eric Qin\IEEEauthorrefmark{1}, Francisco Mu\~noz-Matr\'inez\IEEEauthorrefmark{2}, Robert Guirado\IEEEauthorrefmark{3}, Akshay Jain\IEEEauthorrefmark{4}, Sergi Abadal\IEEEauthorrefmark{3}, \\Jos\'e L. Abell\'an\IEEEauthorrefmark{5}, Manuel E. Acacio\IEEEauthorrefmark{2}, Eduard Alarc\'on\IEEEauthorrefmark{3}, Sivasankaran Rajamanickam\IEEEauthorrefmark{6}, Tushar Krishna\IEEEauthorrefmark{1}}
\\
\IEEEauthorblockA{
\IEEEauthorrefmark{1}Georgia Institute of Technology, \IEEEauthorrefmark{2}Universidad de Murcia, \IEEEauthorrefmark{3}Universitat Politecnica de Catalunya, \IEEEauthorrefmark{4}Neutroon,\\\IEEEauthorrefmark{5}Universidad Cat\'olica de Murcia, \IEEEauthorrefmark{6}Sandia National Laboratories\\ 
Email: \IEEEauthorrefmark{1}\{raveesh.g, ecqin\}@gatech.edu, \IEEEauthorrefmark{1}tushar@ece.gatech.edu, \IEEEauthorrefmark{2}\{francisco.munoz2, meacacio\}@um.es,\\ \IEEEauthorrefmark{3}roberto.guirado.linan@estudiantat.upc.edu, \IEEEauthorrefmark{3}abadal@ac.upc.edu, \IEEEauthorrefmark{3}eduard.alarcon@upc.edu, \IEEEauthorrefmark{4}akshay.jain@neutroon.com,\\ \IEEEauthorrefmark{5}jlabellan@ucam.edu, \IEEEauthorrefmark{6}srajama@sandia.gov}}

\pagestyle{plain}
\maketitle



\begin{abstract}

Graph Neural Networks (GNNs) have garnered a lot of recent interest because of their success in learning representations from graph-structured data across several critical applications in cloud and HPC. Owing to their unique compute and memory characteristics  that come from an interplay between dense and sparse phases of computations, the emergence of reconfigurable dataflow (aka spatial) accelerators offers promise for acceleration by mapping optimized dataflows (i.e., computation order and parallelism) for both phases. The goal of this work is to characterize and understand the design-space of dataflow choices for running GNNs on spatial accelerators in order for mappers or design-space exploration tools to optimize the dataflow based on the workload. Specifically, we propose a taxonomy to describe all possible choices for mapping the dense and sparse phases of GNN inference, spatially and temporally over a spatial accelerator, capturing both the intra-phase dataflow and the inter-phase (pipelined) dataflow. Using this taxonomy, we do deep-dives into the cost and benefits of several dataflows and perform case studies on implications of hardware parameters for dataflows and value of flexibility to support pipelined execution. 

\begin{IEEEkeywords}
Graph Neural Networks, Spatial Accelerators, Dataflows, Pipelined Parallelism
\end{IEEEkeywords}

\end{abstract}
\section{Introduction}
\label{sec:introduction}

Recently, there has been an emergence of several general-purpose programmable spatial accelerators
(aka reconfigurable or flexible dataflow accelerators~\cite{sara_isca2021}) targeting machine learning and HPC. Commercial examples include Cerebras~\cite{cerebras},  Graphcore~\cite{graphcore},
and SambaNova~\cite{sambanova}, and academic prototypes include Plasticine~\cite{plasticine} and MAERI~\cite{kwon2018maeri}.
While low-level 
microarchitectural details vary, 
at a high-level these accelerators 
are comprised of a ``spatial'' array of processing elements (PEs), a private register file (RF) in each PE, a programmable scratchpad-based memory hierarchy, 
and specialized networks-on-chip (NoC) for operand distribution and output collection to/from the PEs. Fig.~\ref{fig:spacc} shows an example.

A key distinguishing feature of  spatial accelerators from conventional CPUs and GPUs is the ability to support dataflow execution~\cite{plasticine}.
In the spatial accelerator domain, the term \textit{dataflow} is used to refer to
the loop transformations 
supported by the accelerators to stage the computation across space (i.e., parallelism) and time (i.e., loop order)~\cite{eyeriss2016isca,kwon2019understanding}.
A ``good'' dataflow is the one 
that can maintain a high utilization via efficient parallelization, 
and minimize data movement through the memory hierarchy via data reuse.
The dataflow along with tile sizes is known as a \textit{mapping}~\cite{kwon2019understanding}. 
While fixed dataflow accelerators~\cite{eyeriss2016isca,kwon2019understanding} encode the dataflow into hardware FSMs,
programmable/flexible accelerators
support diverse mappings which
can be statically configured 
by a compiler.

The space of dataflows for executing dense DNN layers sequentially has been explored heavily~\cite{kwon2019understanding,timeloop,interstellar,eyeriss2016isca}. \hl{These works have proposed \textit{taxonomies} to describe the design-space of DNN dataflows, which have helped the architecture and compiler researchers by providing understanding of reuse{~\cite{kwon2019understanding}},{~\cite{eyeriss2016isca}} and formalizing the design-space of the dataflows for mapping optimization{~\cite{chatarasi2020marvel}},{~\cite{kao2020gamma}}. However, as {Fig.~\ref{fig:venn}} shows, these prior dataflow taxonomies target only a single layer at a time. Moreover, these works only target dense DNNs.
}

Several ML and HPC kernels \hl{(e.g. GCNs{~\cite{kipf2017semisupervised}}, DLRMs{~\cite{naumov2019deep}}, BiCGStab{~\cite{cerebras}} etc.)} employ multiple phases of \textit{sparse-dense} computations offering opportunities for pipelining and reuse across phases. The focus of this work is to characterize and understand the dataflow choices for such \textit{multiphase sparse-dense} kernels, \hl{thus expanding the design-space of the dataflows as {Fig.~\ref{fig:venn}} shows.}
\hl{The design-space of these dataflows is non-trivial due to the interdependence of two phases.} \hl{Moreover, sparsity makes the reuse and compute utilization data-dependent.} \hl{In the given scope of the paper,} this work focuses on understanding the dataflow choices for Graph Neural Networks (GNNs) \hl{inference}, though our analysis can also extend to other 
multiphase computations as well.

\insertFigure{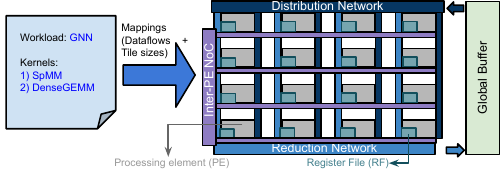}{Spatial Accelerator
}
\insertFigure{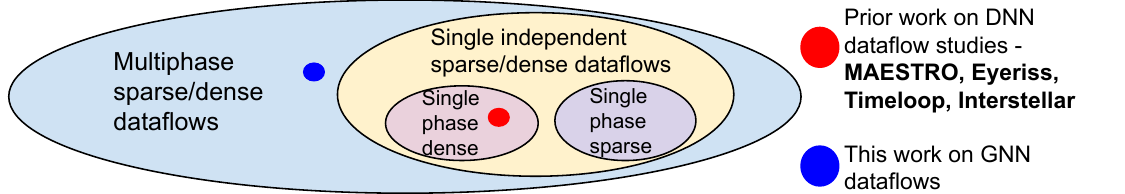}{Map space of our work on GNN dataflow studies
is a superset of that of prior works on DNN dataflow studies - MAESTRO~\cite{kwon2019understanding}, Eyeriss~\cite{eyeriss2016isca}, Timeloop~\cite{timeloop} and Interstellar~\cite{interstellar}.
\vspace{-0.5mm}}


%
%
%
%
GNNs are becoming increasingly popular because of their ability to accurately learn representations from graph-structured data to solve graph/node classification, graph generation, and link prediction problems \cite{wu2019comprehensive}. Example applications include item recommendation \cite{fan2019graph}, molecular feature extraction \cite{duvenaud2015convolutional} and natural language processing \cite{miwa2016end}. 
%
%
GNN inference runtime is dominated by two phases: (1) Aggregation and (2) Combination~\cite{yan2020hygcn}. \textit{Aggregation} is an SpMM computation with irregular, workload dependent accesses of data. \textit{Combination} computations can be cast as GEMMs, similar to 
dense DNNs.

Several recent works on GNN acceleration~\cite{yan2020hygcn, geng2019awb, liang2020engn, Kiningham2020, chen2020rubik, Auten2020}
have demonstrated the challenges with getting high 
performance for GNNs from commodity CPUs, GPUs, and DNN accelerators. 
These challenges arise from (i) extremely high amounts of sparsity in the Aggregation phase (over 99\% as compared to 70-80\% in modern DNNs), and 
(ii) diverse data reuse opportunities within and across these phases.
These works have also demonstrated that 
GNNs offer opportunities for acceleration by crafting specialized \textit{dataflows} for extracting 
reuse both within and across the phases.
Unfortunately, most of these prior works propose 
building specialized GNN accelerator ASICs with 
heavy co-design of a specific GNN dataflow and its hardware microarchitecture, 
which limits their applicability as 
graph datasets and GNN algorithms evolve. In contrast, this is the first work, to the best of our knowledge, that provides a qualitative and quantitative analysis of the \textit{design-space of various inter-phase and intra-phase dataflow strategies for mapping GNNs over flexible spatial accelerators.}

This work makes the following contributions:

\textbf{(i) \textit{\hl{Taxonomy:}}} We propose a succinct taxonomy to classify various inter-phase and intra-phase dataflows to describe the possible pipelining strategies employed by GNN accelerators.
This taxonomy expresses: (1) \textit{Aggregation intra-phase} dataflow, (2) \textit{Combination intra-phase} dataflow, (3) \textit{Inter-phase strategy}, and (4) \textit{phase ordering}.
Rather than targetting 
a specific GNN 
accelerator where the dataflow choices might be limited due to its microarchitecture~\cite{yan2020hygcn,liang2020engn,geng2019awb}, or directly compare two different GNN accelerators where it would become hard to tease apart the performance difference due to dataflow versus microarchitecture, we target a templated \textit{flexible}
spatial accelerator substrate~\cite{plasticine, kwon2018maeri} 
over which any dense/sparse 
dataflow that our taxonomy describes can be run.

\textbf{(ii) \textit{\hl{Qualitative Analysis and Analytical Framework:}}} Using the taxonomy, we explore various factors that affect the runtime and the energy of these dataflow strategies such as  spatial/temporal mapping choices of various dimensions in the intra-phase, the pipelining granularity in the inter-phase, 
and tile sizes. We encode these into a framework called OMEGA.

\textbf{(iii) \textit{\hl{Quantitative Analysis:}}} Using OMEGA, we study the impact of graph properties (such as number of vertices, edges, features) on dataflow choices. We also analyze the impact of hardware parameters, for example, low distribution and reduction bandwidth, and issues of load balancing in pipelining.

\textbf{(iv) \textit{\hl{Architectural Insights:}}} We demonstrate the benefits of reconfigurable dataflow accelerators for multiphase computation as opposed to GNN accelerator ASICs given the interdependence of dataflows.

 \hl{This work explores the design-space of pipelined dataflows with both sparse and dense phases, and proposes the OMEGA framework to model the cost of these dataflows. The insights from this work can extend to other ML and HPC workloads with multiphase kernels{~\cite{cerebras},~\cite{naumov2019deep}}. We envision that the taxonomy will formalize the design-space of multiphase dataflows which will enable better mapping optimizers for ML and HPC workloads. Moreover, the future mapping optimizer can use OMEGA framework as a cost-model for multiphase dataflows. The insights from this work would also help the architects make informed design decisions, thus enabling a richer HW/SW co-design for future ML and HPC accelerators.}

\insertFigure{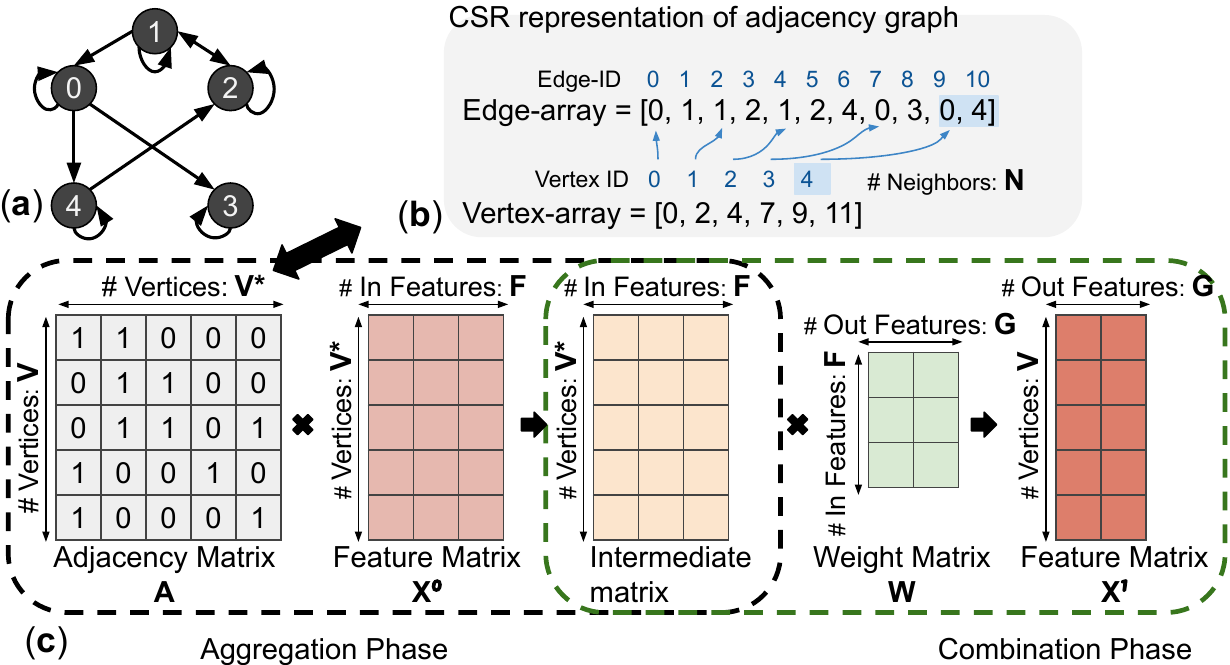}{(a) Example graph, (b) CSR representation of the adjacency matrix, (c) computations. Note: V* can be represented as neighbors(N) if it is represented in CSR format. Also, figure assumes computation order Aggregarion to Combination ((AX\textsuperscript{0})W). For Combination to Aggregation (A(X\textsuperscript{0}W)), intermediate matrix is V$\times$G. 
\vspace{-1mm}} 

\insertWideFigure{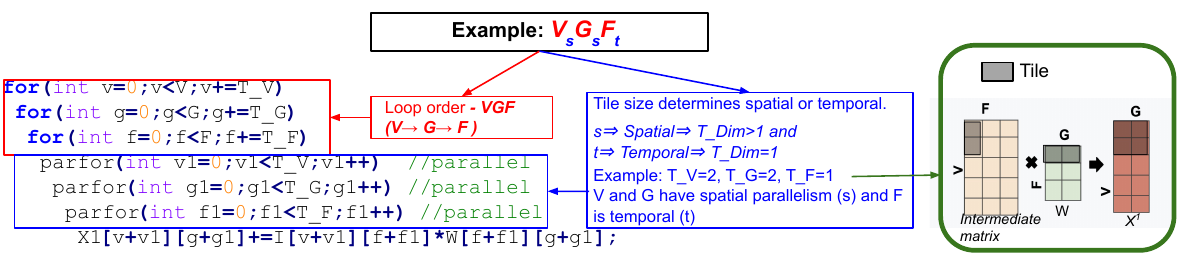}{Example showing compact notation for the loop-nest for Combination phase dataflow \hl{and an example tiling strategy for that dataflow}. See Fig{~\ref{fig:computations}} for dimension and matrix notations.
Similar loop ordering and tiling notation applies to Aggregation.\vspace{-2mm}}

\section{Background and Related Work}
\label{sec:background}
\subsection{GNN Computation}


An input graph is represented by \textit{G(V, E)} with \textit{V} vertices and \textit{E} edges. A graph can also be represented as an adjacency matrix which represents the connectivity of the graph. Matrix representations are common, since dense/sparse matrix multiplication has been the prime target of spatial accelerators. 
Recent works show that the main computation bottlenecks of various GNN algorithms like GCN~\cite{kipf2017semisupervised}, GraphSage~\cite{hamilton2017inductive}, GINConv~\cite{xu2018how} can be broken down into two phases: \textit{Aggregation} and \textit{Combination}~\cite{yan2020hygcn}.
GCNs allow either phase to precede the other while some algorithms like GraphSAGE perform Aggregation before Combination.
%
The Aggregation phase involves the addition of the feature vectors of the neighboring nodes in the graph.
This leads to irregular accesses of data due to the irregular degree distribution in the graphs of interest and, as a result, irregularity in the neighbor locations of a particular vertex in a given dataset. 
The Aggregation phase can be mapped as sparse matrix$\times$dense matrix multiplication (SpMM) with the highly sparse adjacency matrix being the main source of irregular accesses.
The Combination phase is a feature reduction phase.
This phase is dense with regular GEMM computation. 

Fig.~\ref{fig:computations}a) shows an example graph with five vertices and eleven edges, including self loops and Fig.~\ref{fig:computations}c) shows the adjacency matrix representation of the graph. Fig.~\ref{fig:computations}b) is the Compressed Sparse Row (CSR) 
representation of the graph's adjacency matrix. For CSR representation, 
the neighbors of a particular vertex are stored back-to-back. 
Because most graphs are extremely sparse ($>$ 99\% sparsity) \cite{kersting2016benchmark}, CSR is often used as the graph representation \cite{yan2020hygcn, geng2019awb}, and is what we assume in this work.
As labeled in Fig.~\ref{fig:computations}c), \textbf{A} represents the adjacency matrix, \textbf{X} represents the feature matrix, and \textbf{W} represents the weight matrix. Any computation with \textbf{A} is a part of the \textit{Aggregation phase}, and any computation with \textbf{W} is a part of the \textit{Combination phase}. Other variables include: 
\textbf{V} (\# vertices),
\textbf{F} (\# input features),  \textbf{G} (\# output features), and \textbf{N} (\# of neighbors of a vertex).







\subsection{GNN Accelerators}

Several GNN accelerator ASICs have been proposed~\cite{abadal2020computing,liang2020engn,yan2020hygcn,geng2019awb,Kiningham2020,Auten2020,chen2020rubik, gcnax, iscas}, each implementing a 
specific dataflow which is heavily co-designed with the microarchitecture.
The dataflows from these accelerators form a subset of the design-space that we explore.
GCNAX~\cite{gcnax} proposes a flexible GCN accelerator with the ability to choose between some dataflows to optimize for energy.
However, GCNAX primarily targets off-chip dataflows with a small global buffer and 16 PEs, while our work focuses on on-chip dataflow strategies for large programmable spatial accelerators with high parallelism opportunities.


\subsection{Dataflow Analysis}
There have been several efforts to understand and classify dataflows for DNNs such as Timeloop \cite{timeloop}, Interstellar \cite{interstellar}, and MAESTRO~\cite{kwon2019understanding}.
However, as Fig.~\ref{fig:venn} shows, these works focus only on dataflows for single dense layers, forming a subset of dataflows that we study in this work (which includes pipelining and multi-phase sparse-dense dataflows).

\section{GNN Dataflow Taxonomy}
\label{sec:dataflow}

In this section, we discuss the GEMM/SpMM dataflows within an individual phase \textit{(Intra-phase dataflows)} and the overall dataflow with both phases \textit{(Inter-phase dataflows)}. 

\textbf{Notation.} We use the notations from Fig.~\ref{fig:computations} for the matrices and the dimensions
for the rest of the paper.
Fig.~\ref{fig:simplify-mappings}
presents the succinct 
notations we 
use for describing the dataflow of a single phase, 
without writing 
out its full loop-nest. It describes the loop order (i.e., order of temporal loops) and the spatial parallelism choices which are determined by the \textit{tile sizes}.
In this paper, \textit{tile size} (T\textunderscore Dim) represents spatial loop tiling and it refers to the number of elements of a dimension mapped in parallel across PEs. For instance, T\textunderscore F is the number of input features (F) processed in parallel across PEs. Here \textit{s} and \textit{t} in the subscript represent whether a dimension \textit{Dim} is spatial (which means T\textunderscore Dim>1) or temporal (T\textunderscore Dim=1). Also, V and F dimensions are used in both the phases so we use T\textunderscore V\textsubscript{CMB} to describe T\textunderscore V in Combination phase and T\textunderscore V\textsubscript{AGG} for Aggregation.

Table~\ref{tables:gemm} discusses the implication 
of the loop order and the spatial parallelism of the dimensions on 
data movement for three popular 2D GEMM dataflows. Next, we describe various dataflow choices and the taxonomy to describe the intra-phase and the inter-phase dataflows.


\begin{table}[t!]
\scriptsize
\centering
\caption{Example 2D GEMM dataflow choices for combination and their hardware implications. The subscript \textbf{$\mathbf{s}$} on two dimensions represents \textbf{spatial mapping} (i.e., unrolling) of those dimensions across the rows and columns of the accelerator (in a tiled manner), while \textbf{$\mathbf{t}$} represents \textbf{temporal mapping}. "Stationary" means that the matrix is inside the PEs while "streaming means" that the matrix is streamed from GB.}
\label{tables:gemm}

\begin{tabular}{|l|l|l|}
\hline
\textbf{Dataflow} & \textbf{\begin{tabular}[c]{@{}l@{}}Implication of \\ Loop Order\end{tabular}}                                           & \textbf{\begin{tabular}[c]{@{}l@{}}Implication of \\ Spatial Dimensions\end{tabular}}                                                                     \\ \hline
V\textsubscript{s}G\textsubscript{s}F\textsubscript{t}      & \begin{tabular}[c]{@{}l@{}}Output (VG) stationary,\\ Intermediate matrix (VF) \\ and Weights (FG) stream \\ every cycle\end{tabular} & \begin{tabular}[c]{@{}l@{}}Spatial multicast of VF and\\ FG every cycle. Temporal\\ reduction of partial sums\\ within each PE.\end{tabular} \\ \hline
G\textsubscript{s}F\textsubscript{s}V\textsubscript{t}      & \begin{tabular}[c]{@{}l@{}}Weight (FG) stationary,\\ Intermediate matrix (VF) \\streams every cycle\end{tabular}                      & \begin{tabular}[c]{@{}l@{}}Spatial multicast of VF\\ every cycle. Spatial reduction\\ of partial sums across PEs\end{tabular}                    \\ \hline
V\textsubscript{s}F\textsubscript{s}G\textsubscript{t}      & \begin{tabular}[c]{@{}l@{}}Intermediate matrix (VF) \\stationary, Weight (FG) \\streams every cycle\end{tabular}                    & \begin{tabular}[c]{@{}l@{}}Spatial multicast of FG\\ every cycle. Spatial reduction\\ of partial sums across PEs\end{tabular}                   \\ \hline
\end{tabular}
\vspace{-3mm}
\end{table}

\subsection{Intra-phase Dataflow}
\label{sec:intradataflow}

\insertWideFigure{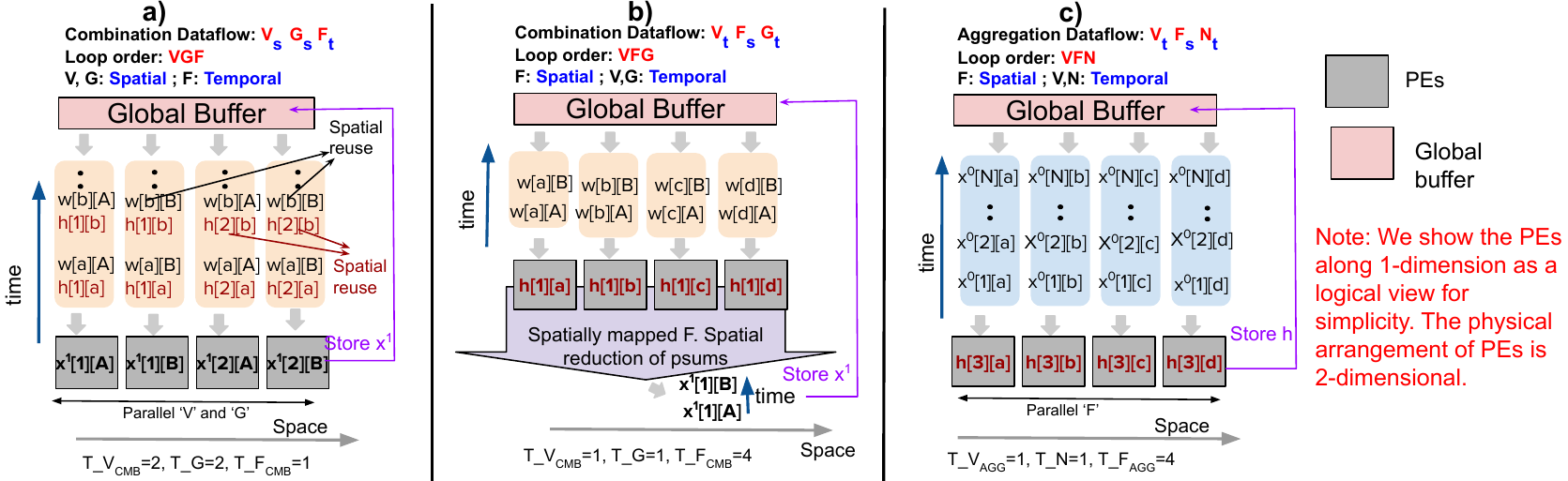}{\hl{Examples of different intra-phase dataflows. a) Combination dataflow: $V_sG_sF_t$, b) Combination dataflow: $V_tF_sG_t$ c) Aggregation dataflow: $V_tF_sN_t$. See {Fig.~\ref{fig:computations}} for matrix and dimension notations. Input feature indices $x^{0}[n=1,2..][f=a,b...]$, Weight indices: $w[f=a,b..][g=A,B..]$, Intermediate indices: $h[v=1,2...][f=a,b..]$, Output feature indices: $x^{1}[v=1,2][g=A,B]$.}}

\subsubsection{\textbf{Combination Dataflow}}
\hl{{Fig.~\ref{fig:intra-phase}}a) and b) 
show examples of Combination dataflows.} In \hl{{Fig.~\ref{fig:intra-phase}a)}}, V$\times$F matrix ($h[v][f]$) and F$\times$G matrix ($w[f][g]$) stream from the buffer into the PEs. The input features vary each cycle, and thus the dimension is temporally mapped and their T\textunderscore F\textsubscript{CMB}=1 due to no spatial parallelism. The other dimensions are spatial since they have parallelism. The tile sizes T\textunderscore V\textsubscript{CMB} and T\textunderscore G are 2, since two vertices `1' and `2' and two output features `A' and `B' are parallelized. Also, `w' and `h' are spatially multicasted in this example. 
This results in V$_s$G$_s$F$_t$.
The reduction is temporal which can be achieved by the accumulators inside the MACs.
As another example, \hl{{Fig.~\ref{fig:intra-phase}b)}} shows a Combination dataflow in which different features are mapped in parallel. V$\times$F matrix is present in the RF. F$\times$G matrix is streamed from the buffer into the PEs. Different partial sums resulting from different features ``a-d'' are reduced spatially which can be done through a linear chain or adder tree~\cite{kwon2019understanding}. This results in V$_t$F$_s$G$_t$ with T\textunderscore V\textsubscript{CMB}=T\textunderscore G=1 and T\textunderscore F\textsubscript{CMB}=4.





\subsubsection{\textbf{Aggregation Dataflow}}
\hl{{Fig.~\ref{fig:intra-phase}c)}} 
illustrates an example of an Aggregation dataflow.
Sparsity gets encoded as N in taxonomy. As Fig.~\ref{fig:computations}b shows, N can be encoded in CSR, in which all of the neighbors are stored back-to-back. 
The example shows V as the outermost loop, and is temporal. 
The next loop contains F which has spatial parallelism ($s$). In the innermost loop, N is temporal, thus temporal reduction is required. This results in V$_t$F$_s$N$_t$. 

\subsection{Inter-phase Dataflow}

\insertFigure{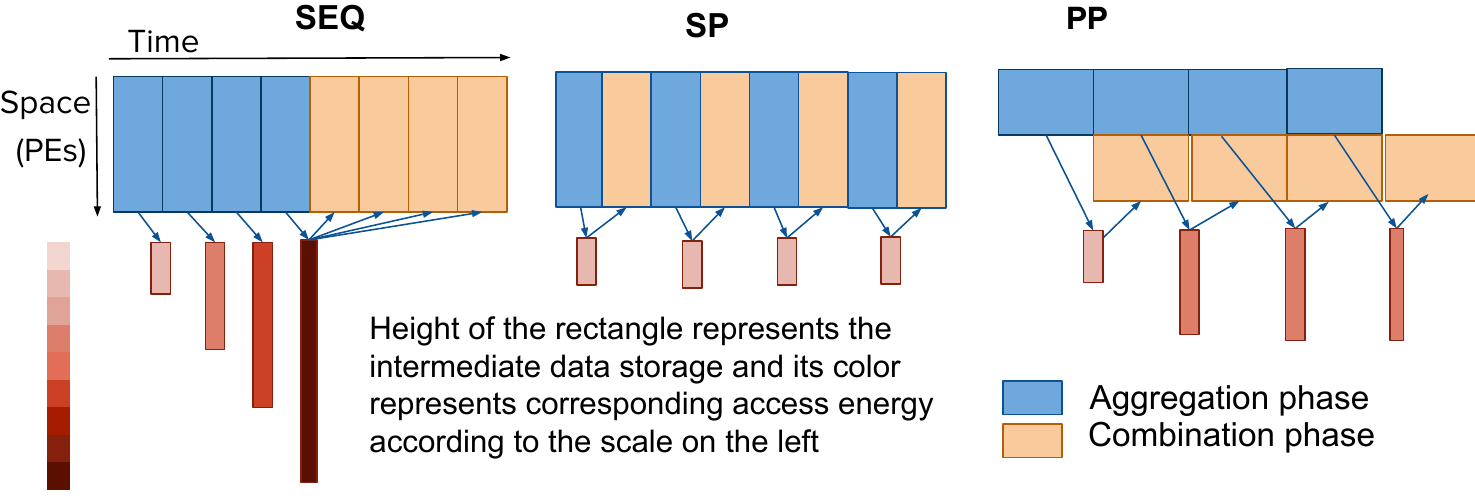}{\hl{Space-Time representation of Inter-phase dataflows along with Intermediate data movement. If the intermediate data exceeds the storage capacity of the on-chip buffers, it needs to move back and forth between memory which adds energy costs.}}

Although each phase can use any intra-phase dataflow, as described above, the dataflow choice for one can affect the other.
This opens up a design-space for 
\textit{inter-phase dataflow}.
This part of the dataflow is important as it determines the number of memory accesses required to move data from one phase to the next.  
{Fig.~\ref{fig:pipeline_figure}} presents the types 
described below. 

\subsubsection{\textbf{Sequential (Seq)}} 

Seq is the simplest inter-phase dataflow where 
the two phases are run sequentially.
All the outputs of the first phase are 
stored in the global buffer (and moved to DRAM if the size of the global buffer is insufficient) 
and then loaded back to the PEs for the next phase.


\subsubsection{\textbf{Sequential Pipeline (SP)}} SP is similar to Seq but it splits the computation of Aggregation and Combination phases into small steps, 
which are then interleaved over time on the 
accelerator's PEs.
As Fig.~\ref{fig:pipeline_figure} shows, it is possible to reduce the data movement between Aggregation and Combination phases depending on the intra-phase dataflows. Specifically, the output data generated by one step can be kept stationary within PEs' local registers
rather than going into a global buffer or DRAM. \hl{{Fig.~\ref{fig:inter-dataflow}b)} shows an example of this. The Aggregation dataflow is same as the dataflow in {Fig.~\ref{fig:intra-phase}}c) and the Combination dataflow is same as the dataflow in {Fig.~\ref{fig:intra-phase}}b) (T\textunderscore F in both phases is 8). In each pipeline step: (a) a part of the intermediate matrix is computed by the Aggregation phase and is stored within the RF of a PE. (b) Weights are streamed over that part of the intermediate matrix to produce a part of the output matrix.} 


\subsubsection{\textbf{Parallel Pipeline (PP)}} PP is when both phases are allocated onto parallel units (i.e., group of PEs) 
within an accelerator at the same time. 
The size of these parallel units can be fixed (e.g., HyGCN~ \cite{yan2020hygcn}) or flexible (e.g., AWB-GCN~ \cite{geng2019awb}),
depending on the flexibility within the accelerator substrate. For PP, it is critical to balance the production and consumption rate to reduce stalls. 
An intermediate ping-pong buffer and NoC.
are needed to send data from one phase to the next. This is similar to the typical workflow for a more general accelerator such as GraphCore~\cite{graphcore}. \hl{{Fig.~\ref{fig:inter-dataflow}a)} shows an example of PP dataflow. In this example, the Aggregation dataflow is same as the dataflow in {Fig.~\ref{fig:intra-phase}}c) and the Combination dataflow is same as the dataflow in {Fig.~\ref{fig:intra-phase}}a). In each pipeline step `n', the left half performs Aggregation and the right half parallelly performs Combination on the part of the intermediate matrix which was produced by Aggregation in the previous step `n-1'.}

\insertWideFigure{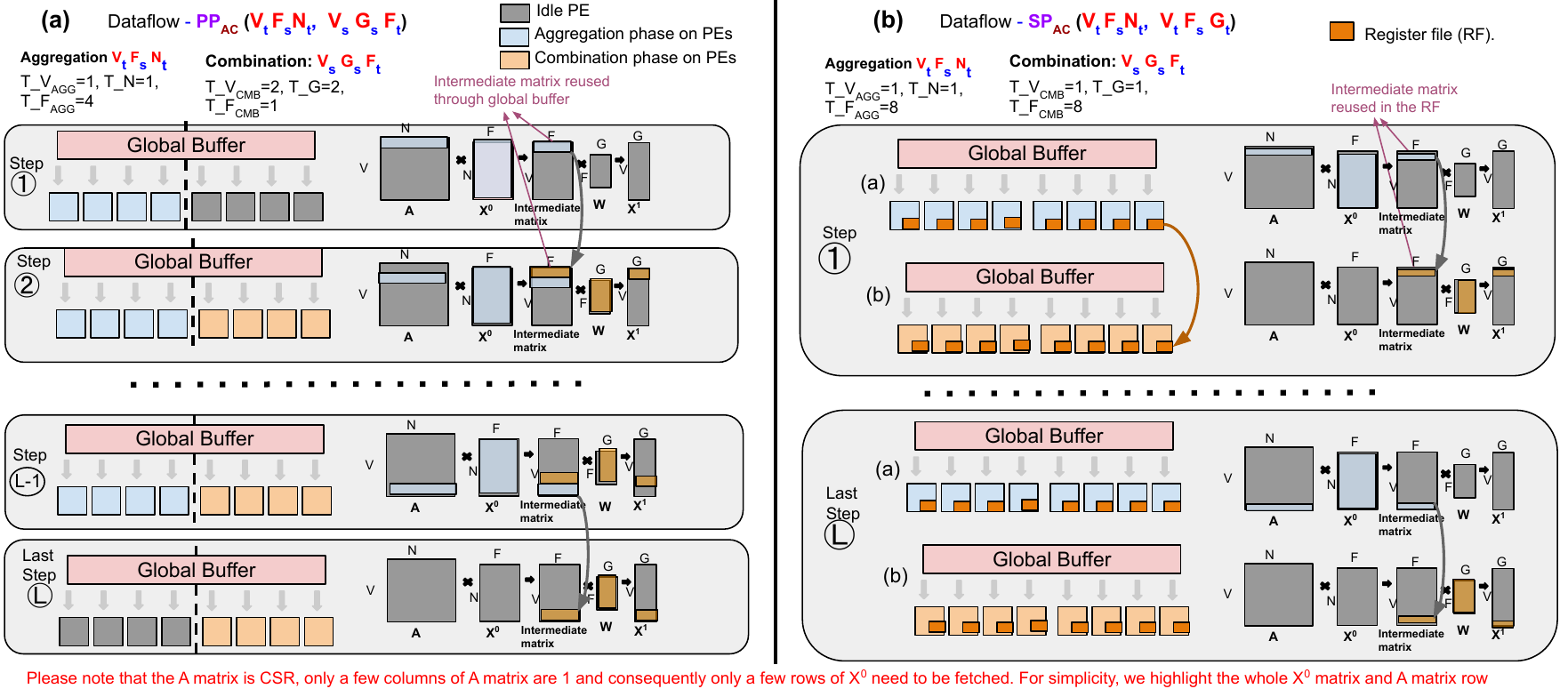}{\hl{Step-by-step execution of inter-phase dataflows.
a) $PP_{AC}(V_{t}F_{s}N_{t}, V_{s}G_{s}F_{t})$ b) $SP_{AC}(V_{t}F_{s}N_{t}, V_{t}F_{s}G_t{t})$. Here, ``step'' refers to duration in which producer writes a part of the intermediate matrix and consumer reads a part of the intermediate matrix. We show the allocation of PEs to the phases and the parts of the matrices computed in each step. See {Fig.~\ref{fig:computations}} for matrix and dimension notations.
Dataflow notation is described in Section{~\ref{sec:taxonomy}}. The intra-phase dataflows are shown in detail in {Fig.~\ref{fig:intra-phase}.}}  }

\vspace{0mm}
\begin{table*}[ht!]
\scriptsize
\centering
\caption{Characterizing the design-space of GNN dataflows. Note that any direction will affect the aggregation and combination dimension variables, but similar concepts apply. Subscripts \textit{s}, \textit{t}, \textit{x} mean spatial, temporal, either spatial or temporal respectively.
}
\begin{tabular}{lllllll}
\toprule
\textbf{Row} & \textbf{Inter Phase}                                                                      & \textbf{\begin{tabular}[c]{@{}l@{}} Order - \textbf{Aggregation}, \\\textbf{Combination}\end{tabular}} & \textbf{\begin{tabular}[c]{@{}l@{}}Intermediate \\Global Buffer\end{tabular}} & \textbf{\begin{tabular}[c]{@{}l@{}}NoC/PE \\ support\end{tabular}}                                                                                                             & \textbf{\begin{tabular}[c]{@{}l@{}}Example \& \\ Order\end{tabular}}      & \textbf{Remarks}                                                                                                                                                                                                                                                                                           \\
\toprule
\textbf{\begin{tabular}[c]{@{}l@{}}1\end{tabular}} & \textbf{\begin{tabular}[c]{@{}l@{}}Sequential\\ (Seq)\end{tabular}}            & ANY-All pairs         & \multicolumn{1}{c}\cmark      & Intra-phase                                                                                                               & \begin{tabular}[c]{@{}l@{}}TPU \cite{tpu-isca} \\ Eyeriss \cite{eyeriss2016isca}\\ (any direction)\end{tabular}     & \begin{tabular}[c]{@{}l@{}}Similar to running one DNN layer at a time.\\
The outputs of one phase get stored in the\\memory and rescheduled back onto the PEs.\end{tabular}                                                                                                         \\ 
\hline
\textbf{\begin{tabular}[c]{@{}l@{}}2\end{tabular}} &
\textbf{\begin{tabular}[c]{@{}l@{}}Sequential\\ Pipeline \\ (SP)\end{tabular}} &  \begin{tabular}[c]{@{}l@{}} AC - V\textsubscript{x}F\textsubscript{x}N\textsubscript{t}, V\textsubscript{x}F\textsubscript{x}G\textsubscript{t}  \\ AC - F\textsubscript{x}V\textsubscript{x}N\textsubscript{t}, F\textsubscript{x}V\textsubscript{x}G\textsubscript{t}  \\ CA - N\textsubscript{x}F\textsubscript{x}V\textsubscript{t}, V\textsubscript{x}G\textsubscript{x}F\textsubscript{t} \\ CA - F\textsubscript{x}N\textsubscript{x}V\textsubscript{t}, G\textsubscript{x}V\textsubscript{x}F\textsubscript{t} \end{tabular}  & \multicolumn{1}{c}\xmark                                                                     & \begin{tabular}[c]{@{}l@{}}Local buffer\\ inside PEs\\ to accumulate\\data\end{tabular} & \begin{tabular}[c]{@{}l@{}}EnGN \cite{liang2020engn}\\ (any direction)\end{tabular}          & \begin{tabular}[c]{@{}l@{}}\textbf{SP-Optimized}: Avoids GB accesses, as output\\ data of one phase is stationary in the PE RF,\\ and can be used as input for the next phase.\\ For agg->cmb, T\textunderscore V\textsubscript{AGG}=T\textunderscore V\textsubscript{CMB} and T\textunderscore F\textsubscript{AGG}\\=T\textunderscore F\textsubscript{CMB} and reduction is temporal as the\\data is always inside in-place buffers. \end{tabular}                                                                   
                                            \\      
                                            \hline
                                            \textbf{\begin{tabular}[c]{@{}l@{}}3\end{tabular}} &
\textbf{}                                                                      & \begin{tabular}[c]{@{}l@{}} Same as rows 4-9  \end{tabular}  & \multicolumn{1}{c}\cmark       & 
\begin{tabular}[c]{@{}l@{}} 
Intermediate outputs\\stored in GB.  \end{tabular}  
&         (any direction)                                                    & \begin{tabular}[c]{@{}l@{}}\textbf{SP-Generic}: There will be a buffer/ setup delay \\ between aggregation and combination phases \\ to remap the output data to a new location.\end{tabular}     

\\                                                
\hline
\textbf{\begin{tabular}[c]{@{}l@{}}4\end{tabular}} &
\textbf{\begin{tabular}[c]{@{}l@{}}Parallel \\ Pipeline\\ (PP)\end{tabular}}   & \begin{tabular}[c]{@{}l@{}} AC - V\textsubscript{x}F\textsubscript{x}N\textsubscript{x}, V\textsubscript{x}F\textsubscript{x}G\textsubscript{x} \\ AC -  F\textsubscript{x}V\textsubscript{x}N\textsubscript{x}, F\textsubscript{x}V\textsubscript{x}G\textsubscript{x} \end{tabular}      & \multicolumn{1}{c}\cmark     & \begin{tabular}[c]{@{}l@{}}NoC connecting Agg\\ and Cmb units to\\intermediate buffer.\end{tabular}                                                          & \begin{tabular}[c]{@{}l@{}}(agg -\textgreater cmb)\end{tabular} & \begin{tabular}[c]{@{}l@{}}\textbf{Element(s) wise granularity}: Element(s) of the \\ intermediate matrix indexed by V,F can be \\ pipelined. 
\end{tabular} \\ 
\hline
\textbf{\begin{tabular}[c]{@{}l@{}}5\end{tabular}} &
& \begin{tabular}[c]{@{}l@{}} AC - V\textsubscript{x}F\textsubscript{x}N\textsubscript{x}, V\textsubscript{x}G\textsubscript{x}F\textsubscript{x}  \\ AC -  V\textsubscript{x}N\textsubscript{x}F\textsubscript{x}, V\textsubscript{x}G\textsubscript{x}F\textsubscript{x} \\ AC -  V\textsubscript{x}N\textsubscript{x}F\textsubscript{x}, V\textsubscript{x}F\textsubscript{x}G\textsubscript{x}  \end{tabular}    & \multicolumn{1}{c}\cmark     & \begin{tabular}[c]{@{}l@{}}NoC connecting Agg\\ and Cmb units to\\intermediate buffer.\end{tabular}                                                             & \begin{tabular}[c]{@{}l@{}}HyGCN \cite{yan2020hygcn} \\ Auten et al. \cite{Auten2020} \\ (agg -\textgreater cmb)\end{tabular} & \begin{tabular}[c]{@{}l@{}} \textbf{Row(s) wise granularity}:Row(s) of the intermediate \\ matrix indexed by V can be pipelined. HyGCN \\ allocates a fixed number of PEs, which may lead \\ to stalls. (combination engine idle waiting). \\HyGCN dataflow-$PP_{AC}(V_x F_s N_t, V_s G_s F_t)$\end{tabular} \\ 
\hline
\textbf{\begin{tabular}[c]{@{}l@{}}6\end{tabular}} &
& \begin{tabular}[c]{@{}l@{}} AC - F\textsubscript{x}V\textsubscript{x}N\textsubscript{x}, F\textsubscript{x}G\textsubscript{x}V\textsubscript{x}  \\ AC -  F\textsubscript{x}N\textsubscript{x}V\textsubscript{x}, F\textsubscript{x}G\textsubscript{x}V\textsubscript{x} \\ AC -  F\textsubscript{x}N\textsubscript{x}V\textsubscript{x}, F\textsubscript{x}V\textsubscript{x}G\textsubscript{x}  \end{tabular}       & \multicolumn{1}{c}\cmark     & \begin{tabular}[c]{@{}l@{}}NoC connecting Agg\\ and Cmb units to\\intermediate buffer.\end{tabular}                                                            & \begin{tabular}[c]{@{}l@{}}(agg -\textgreater cmb)\end{tabular} & \begin{tabular}[c]{@{}l@{}} \textbf{Column(s) wise granularity}: Column(s) of the \\ intermediate matrix indexed by F can be \\ pipelined. \end{tabular} \\ 
\hline
\textbf{\begin{tabular}[c]{@{}l@{}}7\end{tabular}} &
  & \begin{tabular}[c]{@{}l@{}} CA - N\textsubscript{x}F\textsubscript{x}V\textsubscript{t}, V\textsubscript{x}G\textsubscript{x}F\textsubscript{t} \\ CA - F\textsubscript{x}N\textsubscript{x}V\textsubscript{t}, G\textsubscript{x}V\textsubscript{x}F\textsubscript{t} \end{tabular}      & \multicolumn{1}{c}\cmark    & \begin{tabular}[c]{@{}l@{}}NoC connecting Agg\\ and Cmb units to\\intermediate buffer.\end{tabular}                                                            & \begin{tabular}[c]{@{}l@{}}(cmb -\textgreater agg)\end{tabular} & \begin{tabular}[c]{@{}l@{}}\textbf{Element(s) wise granularity}: The order should \\be (NFV, VGF) or (FNV, GVF). V$\times$G matrix \\after Cmb becomes N $\times$ F for Agg. \end{tabular} \\ 
\hline
\textbf{\begin{tabular}[c]{@{}l@{}}8\end{tabular}} &
& \begin{tabular}[c]{@{}l@{}} CA - N\textsubscript{x}V\textsubscript{x}F\textsubscript{x}, V\textsubscript{x}G\textsubscript{x}F\textsubscript{x} \\ CA - N\textsubscript{x}V\textsubscript{x}F\textsubscript{x}, V\textsubscript{x}F\textsubscript{x}G\textsubscript{x}  \\ CA - 
N\textsubscript{x}F\textsubscript{x}V\textsubscript{x}, V\textsubscript{x}F\textsubscript{x}G\textsubscript{x}\end{tabular}    & \multicolumn{1}{c}\cmark     & \begin{tabular}[c]{@{}l@{}}NoC connecting Agg\\ and Cmb units to\\intermediate buffer.\end{tabular}                                                           & \begin{tabular}[c]{@{}l@{}} (cmb -\textgreater agg)\end{tabular} & \begin{tabular}[c]{@{}l@{}} \textbf{Row(s) wise granularity}\end{tabular} \\ 
\hline
\textbf{\begin{tabular}[c]{@{}l@{}}9\end{tabular}} &
& \begin{tabular}[c]{@{}l@{}} CA - F\textsubscript{x}V\textsubscript{x}N\textsubscript{x}, G\textsubscript{x}V\textsubscript{x}F\textsubscript{x} \\ CA - F\textsubscript{x}V\textsubscript{x}N\textsubscript{x}, G\textsubscript{x}F\textsubscript{x}V\textsubscript{x}  \\ CA - 
F\textsubscript{x}N\textsubscript{x}V\textsubscript{x}, G\textsubscript{x}F\textsubscript{x}V\textsubscript{x}\end{tabular}   & \multicolumn{1}{c}\cmark     & \begin{tabular}[c]{@{}l@{}}NoC connecting Agg\\ and Cmb units to\\intermediate buffer.\end{tabular}                                                             & \begin{tabular}[c]{@{}l@{}}AWB-GCN \cite{geng2019awb} \\ (cmb -\textgreater agg)\end{tabular} & \begin{tabular}[c]{@{}l@{}} \textbf{Column(s) wise granularity}: AWB-GCN enables \\a flexible allocation of PEs for different phases \\ to match production and consumption rates.\\ AWB-GCN  dataflow-
$PP_{CA}(F_s N_t V_s, G_t F_t V_s)$
\end{tabular}                                                                                                                           \tabularnewline \bottomrule                                              
\end{tabular}
\label{tables:interleave}
\end{table*}

\subsection{Complete Description of GNN Dataflow}
\label{sec:taxonomy}

Given the dataflow types 
described above, we use the following 
template to describe a complete dataflow:


\textbf{
$\mbox{<Inter>\textsubscript{<order>}(<AggIntra>, <CmbIntra>)}$}

<Inter> represents the inter-phase dataflow, <order> represents the computation order (Aggregation to Combination is AC, Combination to Aggregation is CA). 
<AggIntra> represents the Aggregation phase intra-phase dataflow, and <CmbIntra> represents the Combination phase intra-phase dataflow.








Table~\ref{tables:interleave} uses our taxonomy to enumerate and characterize the space of dataflow
choices for mapping the Aggregation and Combination loops, including the 
hardware structures required for supporting that dataflow.
To keep the table compact, we add a subscript \textit{x} to indicate that the dimension could be spatial or temporal. 
This leads to a total of 6,656 choices
purely from the product of all feasible
loop orders, parallelism choices, and phase order across the three inter-phase choices.
Note that for each dataflow choice, the tile sizes (T\textunderscore Dim) are also parameters which can put the actual number of
possible mappings in the trillions~\cite{timeloop}.
We observe in Table~\ref{tables:interleave} and Section~\ref{sec:analysis} that the loop orders and the tile-sizes of two phases are interdependent for SP and PP dataflows. Thus we formalize the non-trivial space of multiphase GNN dataflows.

For the purposes of analysis, in Table~\ref{tables:interleave} we explicitly list sets of interesting dataflows (some of which existing accelerators have leveraged) with the remarks highlighting their key characteristics.
As an example, the HyGCN~\cite{yan2020hygcn} accelerator's dataflow can be succinctly described as:
$\small PP_{AC}(V_x F_s N_t, V_s G_s F_t)$
In fact, the dataflow shown in
\hl{{Fig.~\ref{fig:inter-dataflow}a)}} is the same as that used by HyGCN~\cite{yan2020hygcn}. However, note that HyGCN microarchitecture employs separate dedicated SIMD and Systolic engines respectively for Aggregation and Combination, along with a dedicated buffer between them for intermediate values. The dataflow is tied to the microarchitecture; in contrast, \hl{{Fig.~\ref{fig:inter-dataflow}a)}} runs the HyGCN dataflow on a programmable spatial accelerator by configuring different sets of PEs to run the Aggregation and Combination dataflows described earlier in Section~\ref{sec:intradataflow} and \hl{staging the intermediate values through the flexible partitions in the programmable scratchpads.}

\textbf{Hardware support for dataflows}.
We 
note that each dataflow 
may require different levels of support from the hardware.
For the intra-phase dataflows, support for multicasts and reductions (either spatially via a  store-and-forward or temporally via a read-modify-write register within PEs) may be needed~\cite{kwon2019understanding}. 
For the inter-phase dataflows, 
Table~\ref{tables:interleave} lists the 
NoC and PE support needed for each dataflow.
We discuss the implications of hardware parameters in detail in Section~\ref{sec:case}.

\insertWideFigure{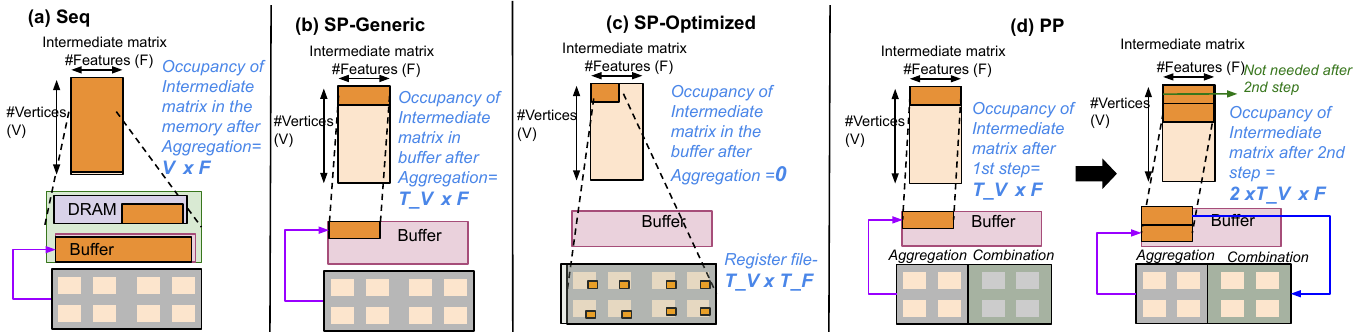}{Intermediate buffering in Inter-Phase dataflows -- AC computation order is assumed. 
} 

\section{Qualitative and Analytical 
Analysis}
\label{sec:analysis}
In this section, we present a qualitative analysis 
of the trade-offs of various inter-phase choices and their runtime and storage implications.
Table~\ref{tables:inter_equations} summarizes the runtime and intermediate buffering requirements for inter-phase dataflows that we derive here. 
We also focus on the intra-phase dataflow implications on inter-phase dataflows where both the phases are interdependent. Thus multiphase dataflow space is non-trivial. For the analysis in this section, we focus on AC computation order, but the same concepts apply to CA. 



\subsection{Sequential Dataflow}
\label{sec:analysis_seq}
Recall that in sequential dataflow, the phases simply run one after the other using any intra-phase dataflow. The overall latency is the sum of latencies of individual phases. The entire intermediate matrix is first written to the memory by the first phase and then read from the memory by the second phase. Hence, as Fig.~\ref{fig:MemHrch}a) shows, the intermediate matrix occupancy in the memory is simply the number of elements of Intermediate matrix which is V$\times$F. Such amount of data cannot be stored on-chip for large graphs, as Fig.~\ref{fig:pipeline_figure} shows, and would incur higher energy than the other inter-phase dataflows.

\subsection{Sequential Pipeline (SP)} 

\label{sec:analysis_sp}
In the Sequential pipeline dataflow, 
a few elements of the first phase are computed and the second phase is applied and this is repeated in an interleaved manner over time.
As Fig.~\ref{fig:pipeline_figure} shows,
this reduces the intermediate matrix footprint and naturally gives an energy advantage over Seq by avoiding expensive memory accesses down in the memory hierarchy (including DRAM).

\textbf{SP-Generic.} This can be considered a na\"{\i}ve implementation for the SP dataflow.
As Fig.~\ref{fig:MemHrch}b) shows, for SP-Generic dataflow, the intermediate data produced in the Aggregation phase is written by the PEs to the global buffer, and then read from there for computing the Combination phase. Thus, the intermediate storage required is equal to the number of elements pipelined which we define as \textit{Pel}. The intermediate data is broken down into granularities which we describe in detail in Section~\ref{sec:analysis_gran}. The feasible loop orders are shown in row 3 of Table~\ref{tables:interleave}, which is equivalent to the PP loop orders shown in rows 4-9. The total runtime of SP-generic is similar to the sequential dataflow, although the energy would be lower by avoiding the transfer of the Aggregation 
outputs to memory.



\insertWideFigure{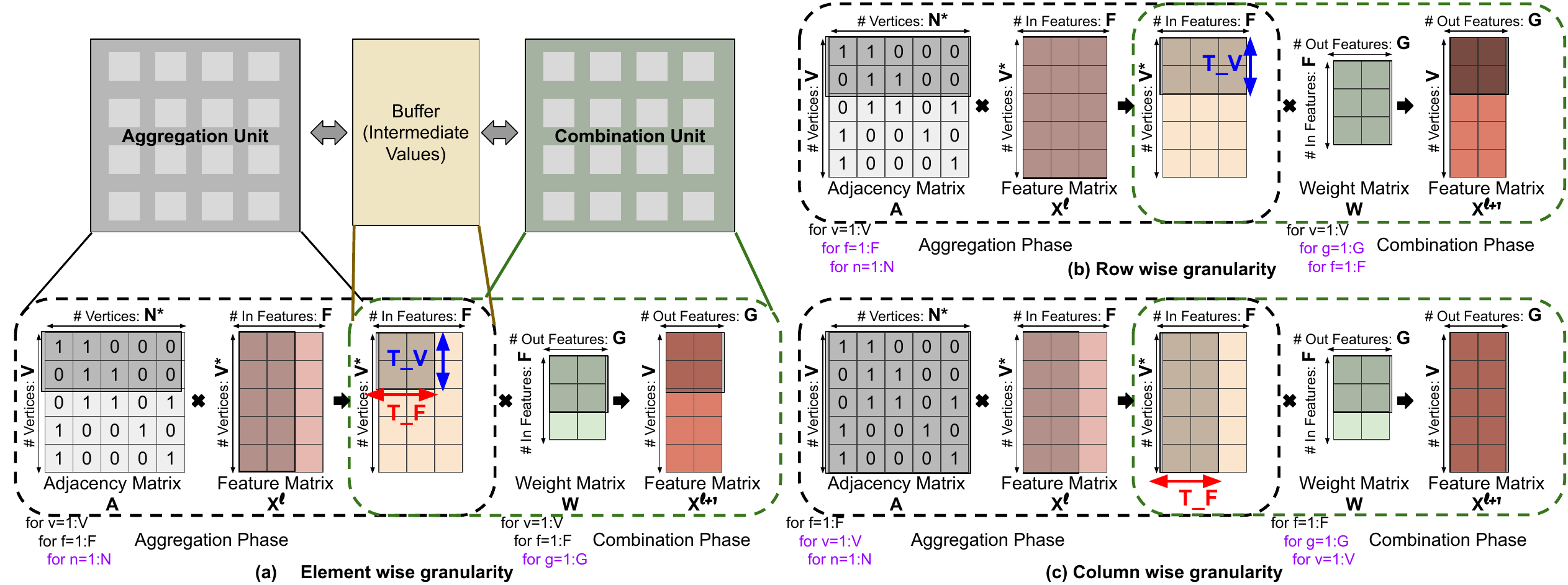}{PP strategies with different granularities: a) Element(s) wise a) Row(s) wise and a) Column(s) wise for the phase order AC.}

\textbf{SP-Optimized.} In order to save both energy and latency,
we can select a subset of intra-phase dataflows for Aggregation and Combination such that the outputs generated by Aggregation can stay local 
within the PEs' registers
and be reused directly 
by the Combination phase. 
This avoids writing this intermediate data up the memory hierarchy to the 
buffer and reading it again.
These dataflows are 
SP\textsubscript{AC}(V\textsubscript{x}F\textsubscript{x}N\textsubscript{t},V\textsubscript{x}F\textsubscript{x}G\textsubscript{t}) and SP\textsubscript{AC}(F\textsubscript{x}V\textsubscript{x}N\textsubscript{t},F\textsubscript{x}V\textsubscript{x}G\textsubscript{t}) for Aggregation to Combination order and are
shown in Row 2 of Table~\ref{tables:interleave}. 
%
In this dataflow, the requirement is that the loop order pair for Aggregation and Combination phases should be (VFN, VFG) or (FVN, FVG). After the first iteration of temporal loop-nest of {V,F} is finished with all neighbors reduced, the accumulated data remains in the MAC units and then dimension G is streamed over it. Moreover, corresponding T\textunderscore Dimensions for both Aggregation and Combination would be same (ie T\textunderscore V\textsubscript{AGG}=T\textunderscore V\textsubscript{CMB} and T\textunderscore F\textsubscript{AGG}=T\textunderscore F\textsubscript{CMB}) since the same intermediate data stored in the PEs by the first phase is processed by the second phase. 
Also, since the data should be available for consumption locally in the processing elements, the reduction must be temporal (T\textunderscore N=1).
The advantage of SP-optimized is reduced buffer accesses and reduced buffer requirement, since the data to be used in the second phase is directly used inside the PEs, as also shown in Fig.~\ref{fig:MemHrch}c). Moreover, it saves the latency and memory read overhead of loading the data into PEs. 
An instance of 
SP-Optimized is used 
by the EnGN~\cite{liang2020engn} accelerator. 
\subsection{Parallel Pipeline (PP)}
\label{sec:analysis_pp}

Parallel pipeline dataflow divides the accelerator into two engines, with one engine feeding the other engine spatially. As Fig.~\ref{fig:MemHrch}d) shows, the intermediate data is broken down into small granularities and the data is processed in a pipelined manner. For example, if granularity is a single row of the intermediate matrix, then the Aggregation phase computes and writes the data corresponding to the n\textsuperscript{th} row and in parallel, the Combination phase reads and processes data corresponding to the (n-1)\textsuperscript{th} row. To facilitate this, intermediate ping-pong buffers are required. As Fig.~\ref{fig:MemHrch}d) shows, the amount of intermediate buffering required is twice the number of elements of the intermediate matrix pipelined. We refer to the number of elements being pipelined as \textit{Pel}. Thus the amount of intermediate storage used is $2\times Pel$, $Pel$ for each phase.
The runtime of one pipeline step is equal to the runtime of the slower phase for producing \textit{Pel} elements. The total runtime is the sum of runtimes of individual steps $sum(max(t_{AGG}, t_{CMB})_{Pel})$.
\subsection{Granularities for SP-Generic and PP}
\label{sec:analysis_gran}
We discuss three types of granularities at which the intermediate matrix can be broken into pipeline steps for PP and SP-Generic dataflows. Rows 4-9 in  Table~\ref{tables:interleave} show all feasible loop orders for all possible granularities and phase orders. 

\textbf{Element.}
Element(s)-wise granularity involves tiles of a few elements 
being processed in a pipelined manner rather than the whole dimensions. Fig.~\ref{fig:example-pipelineparallel}a) shows an example of element(s)-wise granularity PP\textsubscript{AC}(V\textsubscript{x}F\textsubscript{x}N\textsubscript{x}, V\textsubscript{x}F\textsubscript{x}G\textsubscript{x}). 
For each element(s) which is (are) indexed by V and F, the inner-most loop (N) of Aggregation reduces all the neighbors for a given tile of vertices and features to be reduced and, in parallel, the innermost loop of Combination computes G for the previous tile of vertices and features. 
The amount of buffering is 2$\times$\textit{Pel} for PP datalow and \textit{Pel} for SP-Generic. $Pel$ = T\textunderscore V\textsubscript{max}$\times$T\textunderscore F\textsubscript{max}\footnote{If tile size of a dimension (T\textunderscore Dim) is imbalanced, ie (T\textunderscore Dim\textsubscript{AGG} $\neq$T\textunderscore Dim\textsubscript{CMB}), then the T\textunderscore Dim corresponding to Pel would be the LCM. In this work, we only consider mappings where higher tile size is multiple of the lower, which is equivalent to T\textunderscore Dim\textsubscript{max}, otherwise LCM can be large.}. 





\textbf{Row.}
In row(s)-wise granularity, the whole row(s) of the intermediate matrix is (are) considered instead of a few elements in a row. The number of rows that are pipelined is T\textunderscore V\textsubscript{max}. So \textit{Pel} = T\textunderscore V\textsubscript{max}$\times$F.
Fig.~\ref{fig:example-pipelineparallel}b) shows an example loop order (VFN, VGF) for row(s) wise granularity for PP dataflow for phase order Aggregation to Combination. For each set of rows indexed by V, the inner two 
loops F and N compute Aggregation, and in parallel, the two 
innermost loops G and F compute Combination for the previous row(s). An instance of this dataflow is used in HyGCN~\cite{yan2020hygcn}.

\textbf{Column.} 
In column(s)-wise granularity, the whole column(s) of the intermediate matrix is (are) considered instead of a few elements in a column.  The number of columns that are pipelined is T\textunderscore F\textsubscript{max}. So \textit{Pel} = T\textunderscore F\textsubscript{max}$\times$V.
Fig.~\ref{fig:example-pipelineparallel}c) shows an example loop order (FVN, FGV) for column(s) wise granularity for PP dataflow for phase order Aggregation to Combination. Each column of the aggregated matrix is indexed by F while the inner two loops compute the computations corresponding to columns in a pipelined manner.

Table~\ref{tables:inter_equations} summarizes the runtime and buffering requirements for the different inter-phase dataflows discussed before.

\begin{table}[t!]
\scriptsize
  \centering
  \caption{Runtime and buffering requirements for Dataflows} 
  \label{tables:inter_equations}
  \begin{tabular}{|l|l|l|}
    \hline
    \textbf{Inter-phase dataflow} & \textbf{Intermediate Buffering} & \textbf{Runtime}  \\
    
    \hline
    Seq & V$\times$F & t\textsubscript{AGG}+t\textsubscript{CMB}\\
    \hline
    SP-Generic & $Pel$ & t\textsubscript{AGG}+t\textsubscript{CMB}\\
    \hline
    SP-Optimized & 0 & t\textsubscript{AGG}+t\textsubscript{CMB}-t\textsubscript{load}\\
    \hline
    PP-Element & 2$\times$T\textunderscore V\textsubscript{max}$\times$T\textunderscore F\textsubscript{max} & sum(max(t\textsubscript{AGG},t\textsubscript{CMB})\textsubscript{Pel})\\
    \hline
   PP-Row & 2$\times$T\textunderscore V\textsubscript{max}$\times$F & sum(max(t\textsubscript{AGG},t\textsubscript{CMB})\textsubscript{Pel})\\
    \hline
    PP-Column & 2$\times$V$\times$T\textunderscore F\textsubscript{max} & sum(max(t\textsubscript{AGG},t\textsubscript{CMB})\textsubscript{Pel})
    \\
    \hline
  \end{tabular}

\end{table}


%

\section{Quantitative Analysis and Case Studies}

\label{sec:evaluation}
In this section, we do a deep dive into the performance and energy of different GNN dataflows. We also perform case studies on hardware implications.

\subsection{Experimental Methodology
}
\label{sec:methodology}
\subsubsection{Cycle-Accurate Simulation Framework}
In order to carry out a detailed evaluation of various dataflows, we built a cycle accurate simulation framework called OMEGA\footnote{OMEGA codebase - https://github.com/stonne-simulator/omega.} (\underline{\textbf{O}}bserving \underline{\textbf{M}}apping \underline{\textbf{E}}fficiency over \underline{\textbf{G}}NN \underline{\textbf{A}}ccelerators)
around STONNE simulator\cite{stonne}. Fig.~\ref{fig:omega2} shows the overview of the OMEGA framework. STONNE simulator models the flexible accelerators MAERI\cite{kwon2018maeri} and SIGMA\cite{sigma}, and the hardware models have been extensively validated against their RTLs.
\insertFigure{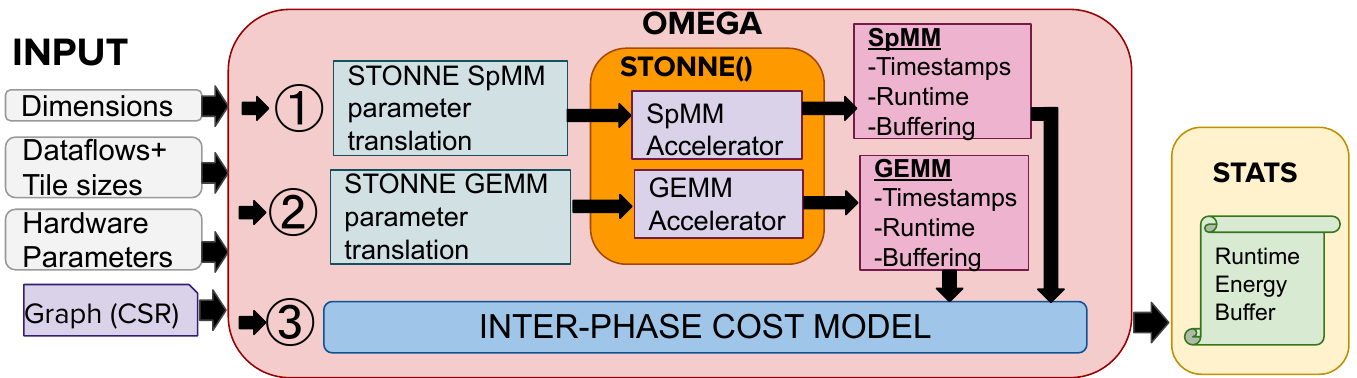}{OMEGA framework toolflow.}
It consists of reconfigurable networks-on-chip 
for operands (e.g., inputs, partial sums and weights) distribution, 
operands multiplication,
and output reduction (e.g., addition) allowing us to study different tile sizes. \hl{The simulator uses a single-cycle distribution network used in the MAERI accelerator.}
The STONNE framework also supports CSR decoding and indexing logic to run SpMM in addition to GEMM to compute both phases. To implement inter-phase dataflows, we built an inter-phase cost model that analytically computes the runtime, buffering, and energy statistics of inter-phase dataflows using statistics from individual phases. \hl{The inter-phase cost model uses an analytical model to compute results for inter-phase dataflows from intra-phase dataflows based on the analysis in Section{~\ref{sec:analysis}}. Some of the example equations are shown in Table{~\ref{tables:inter_equations}}. Some dataflows like PP require timestamps for the portions of outputs computed for both the phases, which are collected at the granularity of $Pel$.}




\subsubsection{Datasets}
We evaluate the GNN dataflows for the
target datasets described in Table~\ref{tables:workload}. These are standard datasets representing workloads from multiple domains like biochemistry, citation networks, and social networks \cite{kersting2016benchmark, yanardag2015deep}.
We evaluate one batch of 64 graphs for graph classification workloads (batch of 32 graphs for RedditBIN) and we evaluate node classification datasets Citeseer and Cora. 
The large graph sets are generally sliced to fit on-chip~\cite{yan2020hygcn,liang2020engn}. Large graph classification datasets can be batched such that the graphs fit on-chip.
For this work, there is sufficient on-chip buffering for a batch of graph classification datasets and for node classification datasets. We characterize on-chip data movement and runtime of the GNN dataflows since our aim is to study the behavior of these dataflows on spatial accelerators. Based on the matrix dimensions and sparsity, we divide the workloads into 3 categories: high number of edges/vertices and relatively low features/vertices \textbf{(HE)}, high features/vertices and relatively lower edges/vertices \textbf{(HF)}, and low edges/vertices and low number of features/vertices \textbf{(LEF)}.

\subsubsection{Evaluation Parameters}\hl{Unless specified otherwise, we assume the number of PEs to be 512 with 64B banked RF in each PE. We also assume sufficient distribution and reduction bandwidth to ensure that the data is received from (or sent to) all the PEs without any stalls. The tile sizes are chosen such that they satisfy the dataflow description in Table{~\ref{tables:eval_config}} and the static utilization\footnote{\hl{Static utilization is $T\_V_{AGG}\times T\_F_{AGG}\times T\_N$ for Aggregation phase and $T\_V_{CMB}\times T\_F_{CMB}\times T\_G$ for Combination phase.}} is nearly 100\% of the PEs. For PP dataflow, unless otherwise mentioned, half the PEs perform Aggregation phase and half the PEs perform Combination phase.}

 

  \vspace{-0.25cm}

\begin{table}[ht!]
\scriptsize
  \centering
  \caption{Datasets information. First part for graph classification, bottom part for node classification -- `*' means that indicator vectors were used in place of features.}
  
  \label{tables:workload}
\begin{tabular}{llllll}
\toprule
\textit{Name}   & \textit{\#Graphs} & \textit{\#Nodes(av)} & \textit{\#Edges(av)} & \textit{\#Features} & \textit{Category} \\ 
\toprule
Mutag
& 188                & 17.93                                                              & 19.79                                                              & 28*        &   LEF       \\
Proteins
& 1113               & 39.06                                                              & 72.82                                                              & 
29 (full)           & LEF        \\
Imdb-bin
& 1000               & 19.77                                                              & 96.53                                                              & 136*       &   HE    \\
Collab
& 5000               & 74.49                                                              & 2457.78                                                            & 492*         &     HE   \\
Reddit-bin
& 2000               & 429.63                                                             & 497.75                                                             & 3782*        &  HF  \\
\hline
Citeseer
& 1                  & 3327                                                               & 9464                                                               & 3703       & HF         \\
Cora
& 1                  & 2708                                                               & 10858                                                              & 1433         & HF       \\

\bottomrule
\end{tabular}
\end{table}

\begin{table}[ht!]
\scriptsize
\vspace{-0.2cm}
  \caption{Dataflow configurations for evaluation.}
  \label{tables:eval_config}
  \begin{tabular}{|l|l|l|}
    \hline
    \textbf{Dataflow Configuration} & \textbf{Name} & \textbf{Distinguishing Property}  \\
    \hline
    Seq\textsubscript{AC}(V\textsubscript{x}F\textsubscript{x}N\textsubscript{t},V\textsubscript{x}G\textsubscript{x}F\textsubscript{x}) & Seq1 & Temporal Aggregation (T\textunderscore N=1)\\
    \hline
    Seq\textsubscript{AC}(V\textsubscript{x}F\textsubscript{x}N\textsubscript{s},V\textsubscript{x}G\textsubscript{x}F\textsubscript{x}) & Seq2 & Spatial Aggregation (T\textunderscore N>1)\\
    \hline
    SP\textsubscript{AC}(V\textsubscript{x}F\textsubscript{s}N\textsubscript{t},V\textsubscript{x}F\textsubscript{s}G\textsubscript{x}) & SP1 & Temporal Aggregation \& high T\textunderscore F \\
    \hline
    SP\textsubscript{AC}(V\textsubscript{s}F\textsubscript{x}N\textsubscript{t},V\textsubscript{s}F\textsubscript{x}G\textsubscript{x}) & SP2 & Temporal Aggregation \& high T\textunderscore V \\
    \hline
     {SP{\textsubscript{AC}}(V{\textsubscript{s}}F{\textsubscript{x}}N{\textsubscript{t}},V{\textsubscript{s}}F{\textsubscript{x}}G{\textsubscript{x}})} & SPhighV & {SP dataflow; extremely high T\textunderscore V} \\
    \hline
    PP\textsubscript{AC}(V\textsubscript{x}F\textsubscript{x}N\textsubscript{t},V\textsubscript{x}G\textsubscript{x}F\textsubscript{x}) & PP1 & Temporal Aggregation \& \\ & & Granularity of lower rows \\
    \hline
    PP\textsubscript{AC}(V\textsubscript{x}F\textsubscript{x}N\textsubscript{s},V\textsubscript{x}G\textsubscript{x}F\textsubscript{x}) & PP2 & Spatial Agg. \& low  granularity \\
    \hline
    PP\textsubscript{AC}(V\textsubscript{x}F\textsubscript{x}N\textsubscript{t},V\textsubscript{s}G\textsubscript{x}F\textsubscript{x}) & PP3 & Temporal Agg. \& high  granularity \\
    \hline
    PP\textsubscript{AC}(V\textsubscript{x}F\textsubscript{x}N\textsubscript{s},V\textsubscript{s}G\textsubscript{x}F\textsubscript{x}) & PP4 & Spatial Agg. \& high  granularity \\
    \hline
   
  \end{tabular}
\end{table}

\insertWideFigure{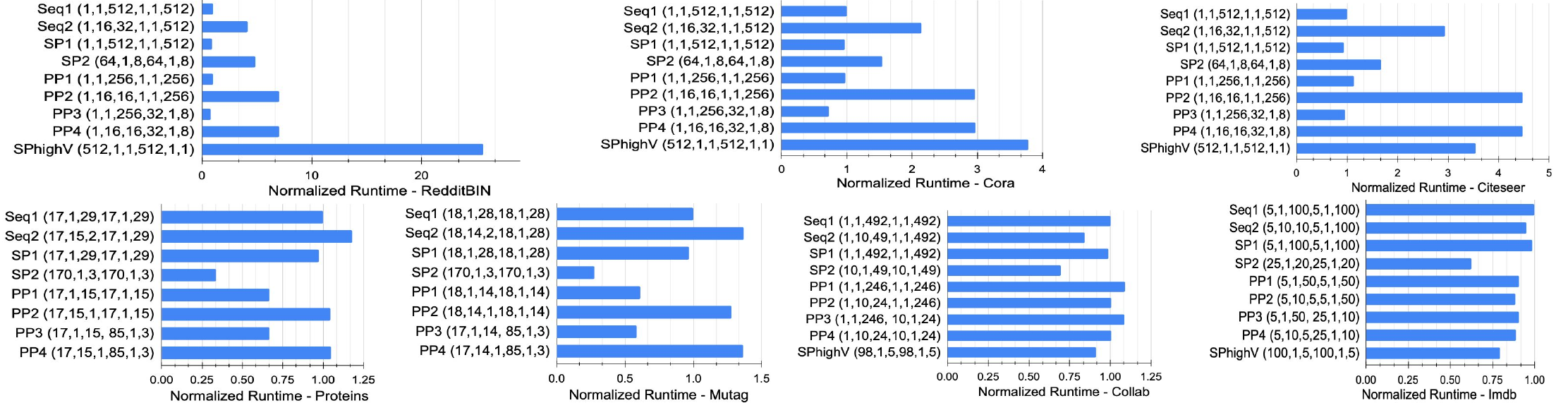}{{Runtimes of Dataflows normalized to Seq1 for GCN algorithm. PE Utilization is close to 100\%. The values in the brackets are tile sizes for each dimension, represented as (T\textunderscore V{\textsubscript{AGG}},T\textunderscore N,T\textunderscore F{\textsubscript{AGG}},T\textunderscore V{\textsubscript{CMB}},T\textunderscore G,T\textunderscore F{\textsubscript{CMB}}).}}

\insertWideFigure{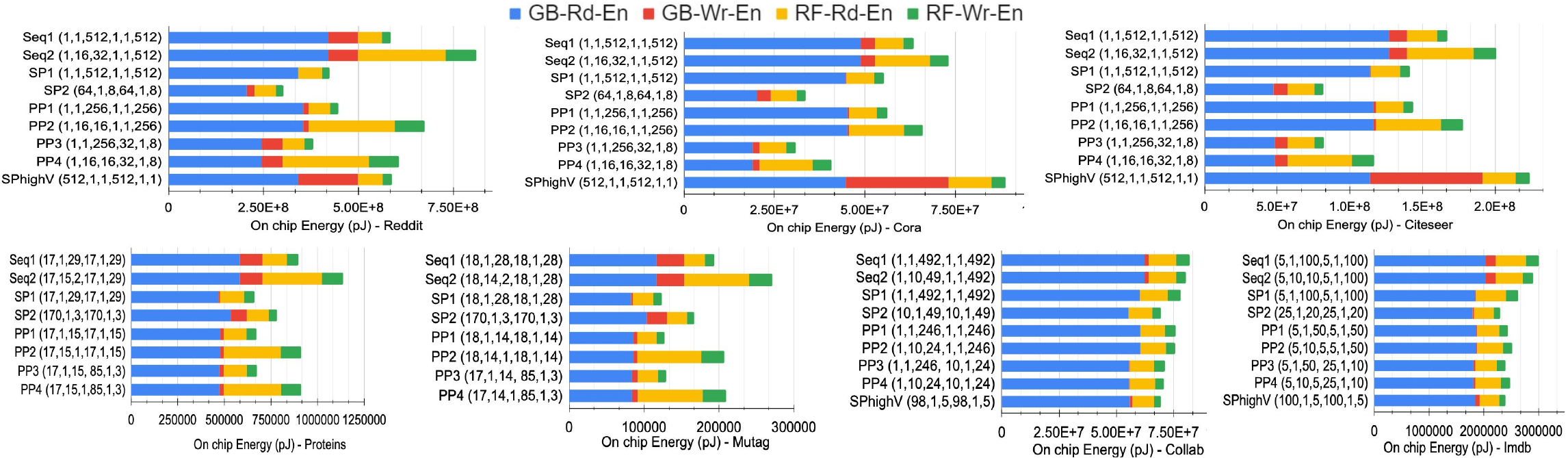}{{On-Chip buffer access energies of dataflows, GB is Global Buffer and RF is local (PE) register file. Size of the Intermediate buffer for PP depends on the pipelining granularity. The values in the brackets are tile sizes for each dimension, represented as (T\textunderscore V{\textsubscript{AGG}},T\textunderscore N,T\textunderscore F{\textsubscript{AGG}},T\textunderscore V{\textsubscript{CMB}},T\textunderscore G,T\textunderscore F{\textsubscript{CMB}}}).}

\insertWideFigure{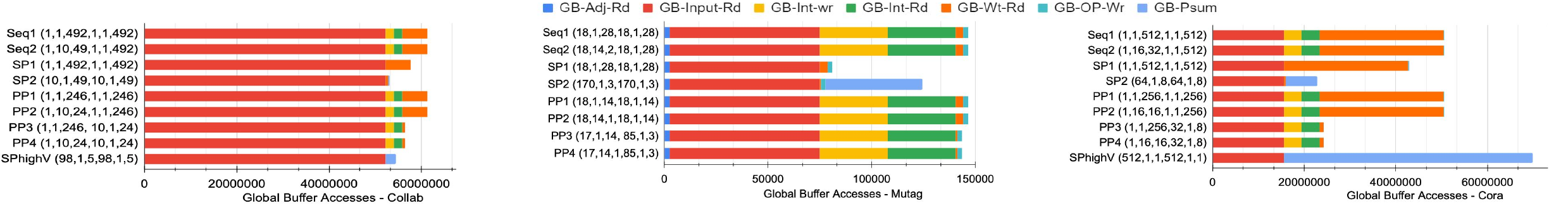}{{Global Buffer breakdown for Mutag and Citeseer, Adj-Adjacency matrix, Inp-Input matrix, Int-Intermediate matrix, Wt-weight matrix and Op-output matrix, Psum-partial sum accesses. The values in the brackets are tile sizes for each dimension, represented as (T\textunderscore V{\textsubscript{AGG}},T\textunderscore N,T\textunderscore F{\textsubscript{AGG}},T\textunderscore V{\textsubscript{CMB}},T\textunderscore G,T\textunderscore F{\textsubscript{CMB}}).}}

\subsection{Comparison of Dataflows}

\label{sec:compare}

We compare the performance and energy of the GNN dataflows. \hl{We evaluate the representative dataflow configurations shown in Table}~\ref{tables:eval_config}.\footnote{\hl{The short names, for example PP4, in the 'Name' column are used to refer to the dataflows in the result charts.} These configurations compare temporal vs spatial Aggregation for Seq and PP dataflows, parallelizing V vs parallelizing F for SP dataflow and granularities of pipelining for PP dataflows. We also introduce SPhighV dataflow to highlight the problem of parallelizing sparse dimensions for all datasets except Proteins and Mutag where T\textunderscore V is already high due to smaller T\textunderscore F limited by F.}

\subsubsection{Performance}
\label{sec:perf}
Fig.~\ref{fig:Performance} shows the runtimes of various dataflows. 
Our observations are as follows:


\insertWideFigure{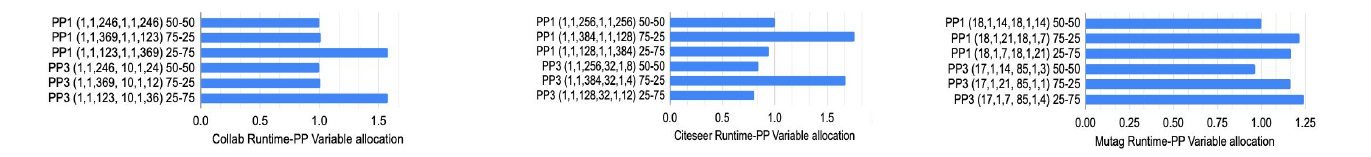}{Runtimes with different PE allocation to phases. Label represents the dataflow and the PE allocation ratio PE\textsubscript{AGG}-PE\textsubscript{CMB}. Runtimes are normalized to 50-50 low granularity. The values in the brackets are tile sizes for each dimension: (T\textunderscore V{\textsubscript{AGG}},T\textunderscore N,T\textunderscore F{\textsubscript{AGG}},T\textunderscore V{\textsubscript{CMB}},T\textunderscore G,T\textunderscore F{\textsubscript{CMB}}).}

\insertWideFigure{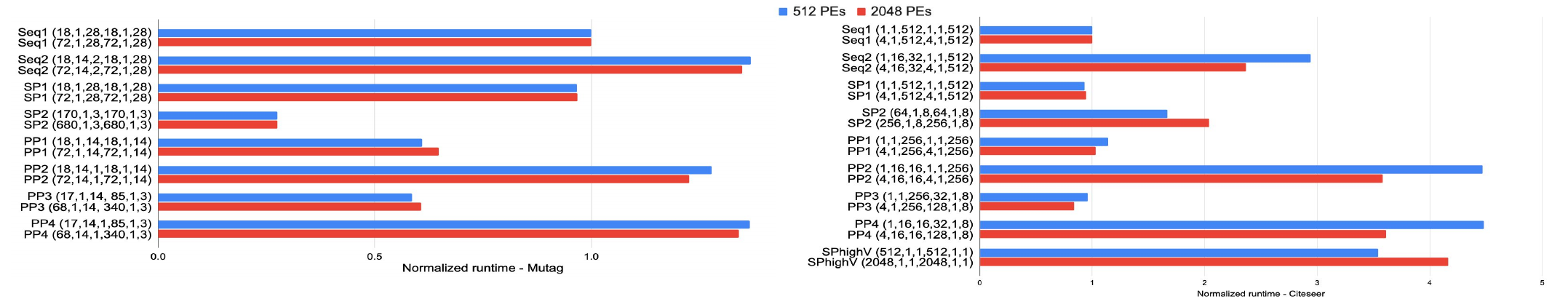}{\hl{Runtime for 2048 and 512 PEs. Runtimes are normalized to that of Seq1 in each case. The values in brackets are tile sizes-(T\textunderscore V{\textsubscript{AGG}},T\textunderscore N,T\textunderscore F{\textsubscript{AGG}},T\textunderscore V{\textsubscript{CMB}},T\textunderscore G,T\textunderscore F{\textsubscript{CMB}}).}}

\insertFigure{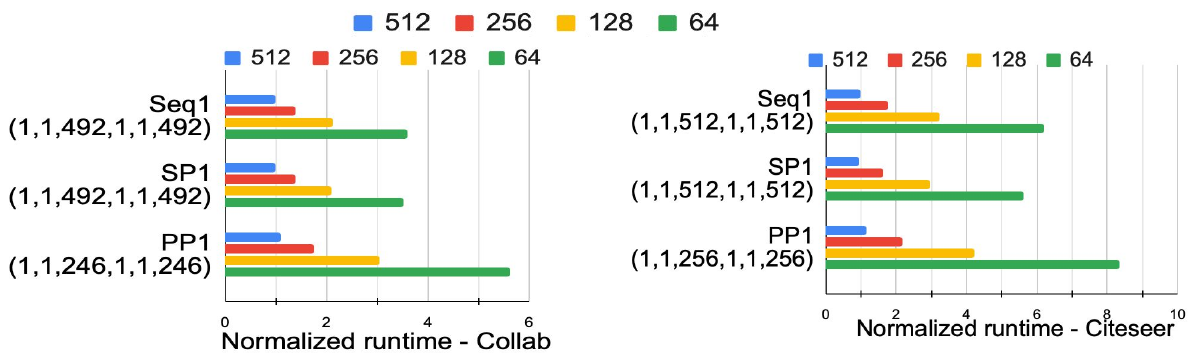}{Runtime normalized to that of Seq1 for 512 elements. Legend represents the number of elements that can be sent to or received from global buffer in parallel. The values in brackets are tile sizes-(T\textunderscore V{\textsubscript{AGG}},T\textunderscore N,T\textunderscore F{\textsubscript{AGG}},T\textunderscore V{\textsubscript{CMB}},T\textunderscore G,T\textunderscore F{\textsubscript{CMB}}).}


\squishlist

\item SP2 (High T\textunderscore V) performs well in most cases since parallelization of the vertices leads to reduced redistribution overhead. However, for HF datasets, extremely high T\textunderscore V can lead to delays since the performance is limited by a dense row (with large number of non-zeros) as demonstrated in SPhighV dataflow. Such dense row is referred to as "evil row" in AWB-GCN~\cite{geng2019awb}.

\item For Collab and Imdb (HE category) spatial Aggregation, in general performs much better than temporal, because they are densely connected. For other datasets, since they are sparse, optimal T\textunderscore N is low. 

\item Mutag and Proteins have great performance despite extremely high T\textunderscore V, since these don't have evil rows. 

\item For the Collab dataset, PP performs worst due to poor load balancing between Aggregation and Combination.

\squishend

\subsubsection{Energy}

\label{sec:energy}
Figure{~\ref{fig:energy-int}} shows the energy consumed by various dataflows across workloads and Fig.~\ref{fig:gbasp} shows global buffer accesses. 
The energy difference comes due to 
differences in the number of accesses 
across the memory hierarchy.
We assume the energy of a global buffer (GB) access to be 1.046pJ (1MB/bank) and the energy of a local PE register file (RF) access to be 0.053pJ based on the energy model from Dally \emph{et al.}{~\cite{ProfDallyTalk}}. For PP dataflow, we assume that there is a separate ping-pong buffer partition for intermediate data and the its size depends on the capacity required based on Table~\ref{tables:inter_equations}. Our observations are as follows:
\squishlist
\item {Energy is dominated by GB reads followed by RF reads.}


\item In Collab (HE category), input GB accesses dominate, in Cora (HF), weight GB accesses dominate. Mutag (LEF) has best percentage of accesses reduced due to reuse.

\item In PP dataflow, we observe lower energy compared to Seq since the energy of memory accesses from smaller intermediate buffer partition is less. SP has no intermediate matrix accesses resulting in low energy.

\item SP2 has low energy but SPhighV has higher energy due to the partial sum accesses due to low T\textunderscore F, especially for HF datasets like Cora. 

\squishend

\subsection{Case Study: Hardware Parameter Implications}

\label{sec:case}
\textit{1) PP: Load balancing.}
In PP dataflow, the delay is decided by the slower phase. Therefore load balancing is critical. Fig.~\ref{fig:ppvar} shows the performance of different allocations of PEs to the phases for different pipeline granularities. Collab has higher density (HE category) hence slow Aggregation, therefore 25-75 performs poorly. For Mutag (LEF category), 50-50 is the best allocation scheme amongst the three allocation schemes
Since, Citeseer is sparse and has high number of features (HF category), the Combination phase is slower, therefore 75-25 allocation performs poorly.

\textit{\hl{2) Scalability of the performance across dataflows.}}
\hl{Fig{~\ref{fig:Performance2048}} shows the performance of the dataflows for an accelerator with 512 and 2048 PEs for Mutag and Citeseer datasets. Tile sizes are chosen to maximize the static utilization. We observe that the runtimes normalized to the Seq1 dataflow are similar in case of 512 and 2048 PEs, especially for dataflows with low runtimes. Therefore, the relative performance of dataflows generalizes for different scales of acceleration.}

\textit{3) Implications of low bandwidth.}
For Section~\ref{sec:compare}, we assumed sufficient on-chip distribution and reduction bandwidth to ensure that the data is received/sent without stalls. However, a low global buffer distribution bandwidth and a low reduction bandwidth lead to stalls affecting the performance. 
Fig.~\ref{fig:bw} shows the performance implications of reducing the bandwidth for different inter-phase dataflows. Runtime reduces with the decrease in the bandwidth and PP dataflow suffers the most since the bandwidth is shared between the two phases.

\subsection{Architectural Insights: Flexibility for Efficient Pipelining}
In this section, we show that flexible accelerators do not only support multiple dataflows but also support efficient pipelining given the interdependence of Aggregation and Combination dataflows in SP and PP. We discuss flexibility features needed in an accelerator to support efficient pipelining.

\textbf{SP-Optimized:} In this dataflow, T\textunderscore V\textsubscript{AGG}=T\textunderscore V\textsubscript{CMB} and T\textunderscore F\textsubscript{AGG}=T\textsubscript{CMB}. Also, in order to retain the intermediate matrix locally in PE, T\textunderscore N=1 (temporal reduction in Aggregation).
For Combination, T\textunderscore F\textsubscript{CMB} determines the spatial reduction. A rigid architecture with only spatial reduction cannot map SP-Optimized. A rigid architecture with only temporal reduction can map only one instance with tile sizes T\textunderscore F=T\textunderscore N=1 which implies that only V is distributed parallelly. This dataflow is SPhighV. We observe from Fig.~\ref{fig:Performance} and~\ref{fig:energy-int} that this dataflow has a huge runtime due to performance being limited by the sparse rows. It also has a huge energy value due to the overhead of writing and reading partial sums. Therefore, \textit{configurability of tile sizes} is essential given the interdependence of phases.

\textbf{PP dataflow}: 
We observed in Fig.~\ref{fig:ppvar}, how critical load balancing is, and that the optimal allocation to a phase can change depending on workload sparsity and dimensions. Using a rigid substrate with two distinct sub-accelerators like HyGCN~\cite{yan2020hygcn} causes load imbalance for certain workloads. Load balancing also requires tile sizes such that stalls are minimized due to the slower phase. Therefore \textit{flexible resource allocation} and \textit{configurability of tile sizes} are essential for mapping PP efficiently.


\textbf{Flexibility features}: Mapping SP and PP dataflows efficiently requires \textit{configurable tile sizes} and \textit{flexible resource allocation}. Moreover as Section~\ref{sec:compare} shows, \textit{flexibility to choose from SP and PP} according to the workload leads to an optimal dataflow. These features add up to a programmable spatial accelerator~\cite{cerebras,sambanova,graphcore,kwon2018maeri,plasticine}. However, there is no additional cost for a programmable spatial accelerator running pipelined dataflows compared to running single phase dataflows. Therefore flexibility provides more benefit per cost for mapping multiphase kernels than independent kernels.

\subsection{Summary of Key Results}

\textbf{\textit{Best Performance}}: For HF workloads, PP3 dataflow is the best, while for other datasets SP2 performs the best, although to achieve the best performance, T\textunderscore V should neither be too high, nor be too low (Fig.~\ref{fig:Performance}).

\textbf{\textit{Best Energy}}: For HF workloads, PP3 and SP2 have the best energies. For HE workloads, SP2 has the best energy. However, energy saving by pipelining is not significant (Fig.~\ref{fig:energy-int}). For LEF workloads, SP1 and PP1 have the best energies.


\textbf{\textit{Cost of pipelining}}: SP-Optimized dataflow can lead to a huge partial sum overhead specially for HF datasets (Section~\ref{sec:energy}). PP dataflow suffers from load balancing problems and is highly sensitive to bandwidth changes (Section~\ref{sec:case}).

\textbf{\textit{Flexibility}}: Flexible accelerators enable choices in dataflow and is also beneficial for efficient SP and PP execution.





\section{Discussions and Future work}
\label{sec:extensions}
\textbf{\textit{Scope of the Dataflow Taxonomy.}}
The current taxonomy captures the intra-phase dataflows and the inter-phase dataflows. 
However, our taxonomy does not capture the order of nodes, graph partitioning and optimizations
such as load balancing~\cite{geng2019awb},  computation elimination via 
memoizing~\cite{chen2020rubik,jia2019redundancy} and requires an extension to capture these.

\textbf{\textit{Application of Taxonomy to Other Kernels beyond GNNs.}} Though this work focuses on GNNs,
the taxonomy and inter-phase analysis proposed in Sections~\ref{sec:dataflow} and~\ref{sec:analysis} can be generalized to dataflows for multiphase computations (GEMM-GEMM/GEMM-SpMM/SpMM-SPMM). One immediate example is Deep Learning Recommendation Models\cite{naumov2019deep} that is built of an SpMM and a DenseGEMM in parallel followed by concatenation followed by a DenseGEMM. 

\textbf{\textit{Mapping Optimizer.}}
In this work, we performed case studies on select dataflows to demonstrate interesting insights.
A mapping optimizer can be built on top of the OMEGA framework, which automatically searches the search space of the dataflows. There are existing mapping optimizers for DNN accelerators~\cite{interstellar, chatarasi2020marvel,kao2020gamma}. Since GNN dataflows add an additional inter-phase optimizations knob, dataflow search is important.

\section{Conclusion}
\label{sec:conclusion}
With the increasing popularity of multiphase workloads like GNNs, dataflow strategies
to map them on accelerators and extract reuse both across and within the phases are crucial.
While there has been prior work on dataflow exploration for dense DNNs, GNNs are a wider generalization of DNNs since they consist of sparse and dense phases. We capture the design space of GNN dataflows in a succinct taxonomy template. Using this taxonomy, we contrast various GNN dataflows and perform various case studies on the pipelined dataflows.

\hl{Rather than targeting a specific ASIC, we explore the design-space of GNN dataflows on a programmable spatial accelerator. We observe that the choice of the dataflow can be influenced by the workload sparsity and dimensions. We also observe various costs and benefits of pipelining. We demonstrate that a flexible accelerator is a more efficient substrate for pipelining due to the interdependence and sensitivity to load balancing in the pipelined dataflows.}



\section*{Acknowledgements}
We thank Prasanth Chatarasi for his insightful feedback. Support for this work was provided through the ARIAA co-design center funded by the U.S. Department of Energy (DOE) Office of Science, Advanced Scientific Computing Research program. Sandia National Laboratories is a multimission laboratory managed and operated by National Technology and Engineering Solutions of Sandia, LLC., a  wholly owned subsidiary of Honeywell International, Inc., for the U.S. Department of Energy's National Nuclear Security Administration under contract DE-NA-0003525. Parts of this work were supported through a fellowship by NEC Laboratories Europe, Project grant PID2020-112827GB-I00 funded by MCIN/AEI/ 10.13039/501100011033, RTI2018-098156-B-C53 (MCIU/AEI/FEDER,UE) and grant 20749/FPI/18 from Fundación Séneca.

\bibliographystyle{IEEEtran}
\bibliography{refs_IPDPS}

\end{document}